\documentclass[english, aps, prb, reprint, amsmath, amssymb]{revtex4-2}
\usepackage[T1]{fontenc}
\usepackage[latin9]{inputenc}
\usepackage{amssymb}
\usepackage{siunitx}
\usepackage{graphicx}
\usepackage{xcolor}
\graphicspath{{Figures/}}
\usepackage{babel}
\usepackage[normalem]{ulem}
\usepackage[colorlinks=true,citecolor={blue!50!black},linkcolor=black,urlcolor=black]{hyperref}

\begin{document}

\title{Towards a $\cos(2\varphi)$ Josephson element using aluminum junctions with well-transmitted channels}

\author{J. Griesmar$^{1}$}
\author{H. Riechert$^{1}$}
\author{M. Hantute$^{1}$}
\author{A. Peugeot$^{1,2}$}
\author{S. Annabi$^{1}$}
\author{\c C. \"O. Girit$^{3}$}
\author{G. O. Steffensen$^{4}$}
\author{A. L. Yeyati$^{4}$}
\author{E. Arrighi$^{1}$}
\author{L. Bretheau$^{1*}$}
\author{J.-D. Pillet$^{1}$}

\altaffiliation{These authors supervised equally this work.
\newline
landry.bretheau@polytechnique.edu 
\newline
jean-damien.pillet@polytechnique.edu}

\affiliation{$^{1}$Laboratoire de Physique de la Mati\`ere condens\'ee, CNRS, Ecole Polytechnique, Institut Polytechnique de Paris, 91120 Palaiseau, France}
\affiliation{$^{2}$ Ecole Normale Sup\'erieure de Lyon, CNRS, Laboratoire de Physique, F-69342 Lyon, France}
\affiliation{$^{3}$ Quantronics Group, Universit\'e Paris Saclay, CEA, CNRS, SPEC, 91191 Gif-sur-Yvette, France}
\affiliation{$^{4}$ Departamento de F\'isica Te\'orica de la Materia Condensada and Condensed Matter Physics Center (IFIMAC),
Universidad Aut\'onoma de Madrid, Madrid, Spain}

\begin{abstract}
We introduce a novel method for fabricating all-aluminum Josephson junctions with highly transmitted conduction channels. Such properties are typically associated with structures requiring intricate fabrication processes, such as atomic contacts or hybrid junctions based on semiconducting nanowires and 2D materials. In contrast, our approach relies solely on standard nanofabrication techniques. The resulting devices exhibit a key signature of high-transmission junctions - Multiple Andreev Reflections (MAR) - in their current-voltage characteristics. Furthermore, we propose a straightforward superconducting circuit design based on these junctions, enabling the implementation of a parity-protected qubit.
\end{abstract}

\maketitle

\section{Introduction}

The tunnel Josephson junction is the cornerstone of superconducting devices. Its Josephson potential $-E_J \cos(\varphi)$ provides the nonlinearity required to create quantum technologies, such as quantum limited amplifiers~\cite{yurke_observation_1989, bergeal_phase-preserving_2010, macklin_nearquantum-limited_2015, aumentado_superconducting_2020} and qubits~\cite{kjaergaard_superconducting_2020, blais_circuit_2021, martinis_quantum_2020}, including the building block of some of the largest quantum processors: the transmon~\cite{koch_charge-insensitive_2007}. The cosine of the superconducting phase difference $\varphi$ arises from the passage of individual Cooper pairs through the tunnel barrier. The number of conduction channels and their transmission $\tau\ll 1$ determines the Josephson energy $E_J$. Higher harmonics $\cos(n\varphi)$ also emerge due to simultaneous tunneling of $n$ Cooper pairs~\cite{senkpiel_single_2020}, but are heavily suppressed as they scale as $\tau^n$. They become significant in junctions with well-transmitted conduction channels of larger transmission $\tau\sim 1$, and can be harnessed to engineer novel quantum technologies such as qubits resilient to decoherence~\cite{ioffe_possible_2002, doucot_pairing_2002, gladchenko_superconducting_2009, bell_protected_2014, smith_superconducting_2020, larsen_parity-protected_2020, schrade_protected_2022, hays_non-degenerate_2025}, three-wave mixing elements \cite{schrade_dissipationless_2024}, or Josephson diodes~\cite{souto_josephson_2022, fominov_asymmetric_2022, valentini_parity-conserving_2024, ciaccia_gate-tunable_2023}.

Josephson junctions with well-transmitted conduction channels are obtained in hybrid architecture S-X-S, where the tunnel barrier is replaced by a quantum conductor X such as a carbon nanotube~\cite{cleuziou_carbon_2006, jarillo-herrero_quantum_2006, annabi_josephson_2024}, a semiconducting nanowire~\cite{goffman_conduction_2017, spanton_currentphase_2017}, graphene~\cite{calado_ballistic_2015, borzenets_ballistic_2016, lee_graphene-based_2020}, or a semiconducting two-dimensional electron gas~\cite{shabani_two-dimensional_2016, hendrickx_ballistic_2019, leblanc_gate-_2025}. The nanofabrication of these hybrid junctions, especially in large quantities, remains a significant challenge as these materials are delicate to handle and the success rate is moderate. On top of that, the interface quality between semiconductors and superconductors is often poor, leading to the formation of parasitic subgap-states~\cite{takei_soft_2013, chang_hard_2015}. Eventually, these devices, being tunable with a gate voltage, are also highly sensitive to voltage noise on the gate. All these imperfections are potential sources of dissipation and decoherence that limit the performances of quantum devices. Consequently, despite their potential for novel quantum technologies, hybrid junctions have not yet gained the same prominence as tunnel junctions, which are more reliable, easier to fabricate, and more stable.

In this work, we introduce a nanofabrication technique for obtaining all-aluminum Josephson junctions with well-transmitted conduction channels, dubbed pinhole Josephson junctions (pinJJs). They derive from tunnel junctions, but with an intentionally inhomogeneous barrier obtained on the roughened edge of a superconducting wire. Although we do not control the exact number of conduction channels nor their transparency, we typically obtain 5-100
channels of which several have high transparency. These channels, which carry most of the supercurrent, are referred to as pinholes. The nanofabrication of pinJJs is achieved using optical lithography and metallic e-beam evaporation with a recipe accessible in standard cleanrooms. We detect the pinholes by measuring the sub-gap current caused by multiple Andreev reflections (MAR)~\cite{averin_ac_1995, bratus_theory_1995, cuevas_hamiltonian_1996}, an electrical signal that only appears for Josephson junctions with well-transmitted conduction channels. By analyzing this signal, we evaluate the number of pinholes as well as their transmission distribution. Our observations confirm their potential for generating higher harmonics $\cos(n\varphi)$. To build on that, we simulate how a combination of four of our junctions in parallel can be used to realize a functional $\cos(2\varphi)$ element, potentially useful to engineer a protected qubit.

\section{Nanofabrication of Pinhole Josephson Junctions (pinJJs)}

In standard tunnel Josephson junctions, the supercurrent is carried by millions of conduction channels with very low transmission, typically  $\tau\sim 10^{-6}$. However, irregularities in the tunnel barrier might lead to the formation of individual channels with transmission close to $\sim 0.1$~\cite{bayros_influence_2024}, which is directly visible in the properties of some transmon qubits~\cite{willsch_observation_2024}. The recipe to fabricate pinJJs is based on this idea: create an intentionnaly inhomogenous barrier with weak points providing a moderate number of channels with large transmission.

\begin{figure}
\includegraphics[width=1\columnwidth]{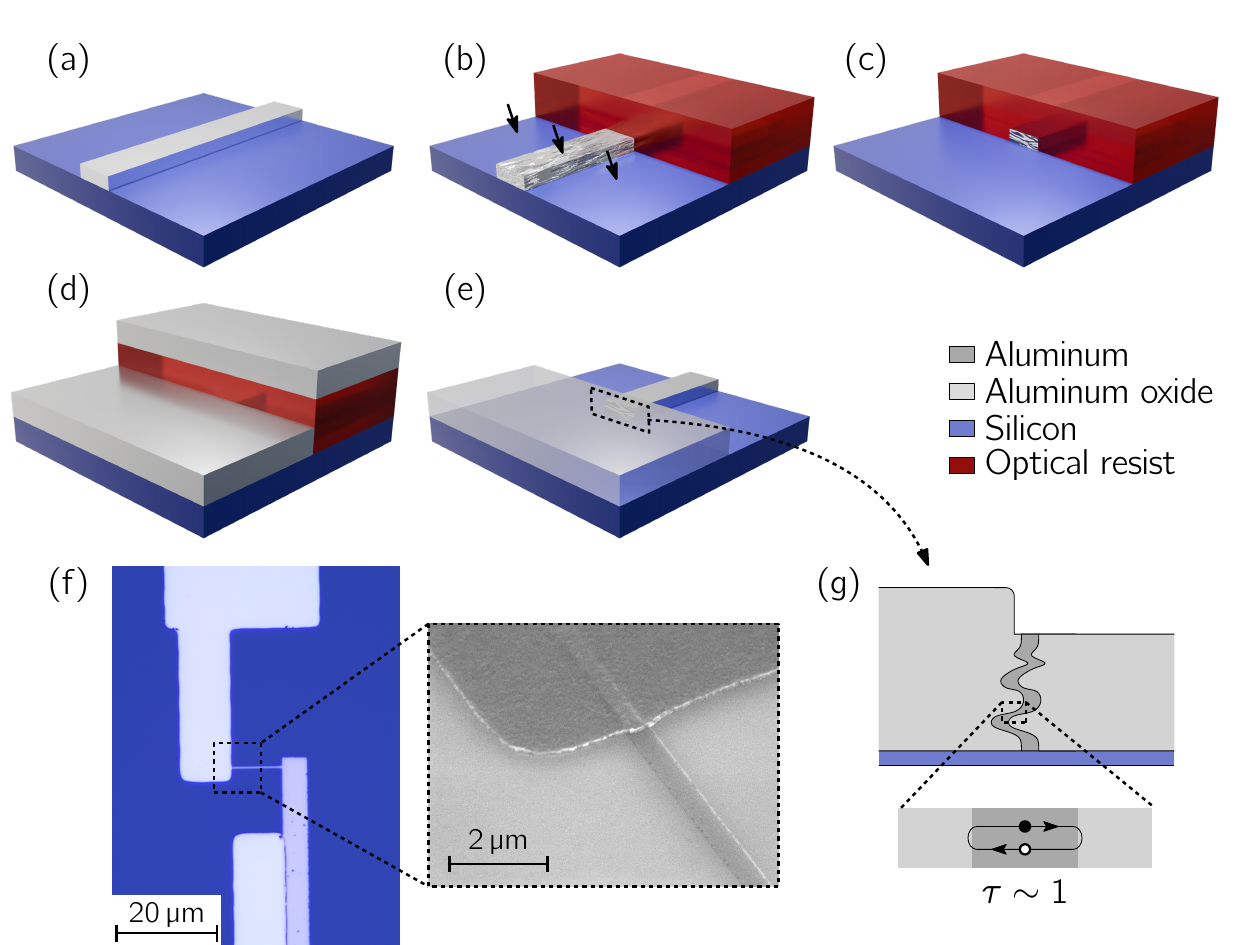}
\caption{\label{fig:fab}Fabrication process.
(a) Schematic view of the device after the first aluminum evaporation.
(b) Argon milling of the oxidized aluminum layer. The arrows symbolize the direction of the milling.
(c) Oxidation of the remaining aluminum surface. Wiggly lines represent the roughness of the etched aluminum surface.
(d) Evaporation of the second aluminum layer.
(e) View of the device after final resist removal.
(f) Optical image of a fabricated device. Inset: SEM picture of the overlap region.
(g) Side-view sketch of the resulting junction with exaggerated roughness. Close-up view: representation of a conduction channel.
}
\end{figure}

To nanofabricate pinJJs, we ion-mill a thin aluminum wire, through a resist mask, across its entire thickness. This exposes an edge with a roughened surface (Fig. \ref{fig:fab}a and b). The surface is then oxidized with a controlled partial pressure of oxygen $P_\mathrm{ox}$ in order to create a tunnel barrier with the desired transparency (Fig. \ref{fig:fab}c). We complete the junction by depositing a second thicker electrode of aluminum, which connects the first wire exclusively over its edge since its top surface is entirely covered by the resist mask (Fig. \ref{fig:fab}d). This process thus forms a tunnel junction on the side of the wire, rather than on the surface like standard tunnel junctions, which is reminiscent of the side contact technique used to connect hexagonal boron nitride encapsulated graphene~\cite{dean_boron_2010, wang_one-dimensional_2013}. From the etching step to the final deposition, everything is done in the same chamber of the e-beam evaporator, including the oxidation, without exposition to air (details of nanofabrication in Appendix \ref{detailfab}).

We first characterize the pinJJs by measuring their electrical resistance at room temperature $R_J$. For $P_\mathrm{ox}=0$, we mostly obtain electrical shorts, while for $P_\mathrm{ox}$ approaching atmoshperic pressure $R_J$ can reach a few M$\Omega$. For our needs, we aim for typical resistance values in the range of a few k$\Omega$, which we achieve with $P_\mathrm{ox}\sim1-10$~mbar and two minutes of oxidation. This might correspond to a handful of well-transmitted channels or rather millions of tunnel ones, but these two scenarios cannot be discriminated with room temperature measurements. Indeed, a perfectly transmitted channel has the same resistance $R_Q=12.9$~k$\Omega$ - the quantum of resistance - as a million channels with transmission $\tau=10^{-6}$. Appendix~\ref{sec:RT_R} shows the typical room temperature resistance distributions obtained for two different oxidations.

\section{Transmissions distribution from Multiple Andreev Reflections}

We now turn to low-temperature measurements to fully characterize the pinJJs by exploiting the phenomenon of multiple Andreev reflections (MAR)~\cite{averin_ac_1995, bratus_theory_1995, cuevas_hamiltonian_1996}. Under a voltage bias $V$, a dissipative current flows through a JJ with an amplitude that is highly dependent on the weak link transmissions ${\tau_i}$. This quasiparticle current arises from successive Andreev reflections at the interfaces between the weak link and the superconducting electrodes. When $\frac{2 \Delta}{n}\le |eV| < \frac{2 \Delta}{n-1}$, the $n$-th order MAR process is dominant ($n\geq2$ being an integer), which involves $n-1$ Andreev reflections and the transfer of a charge $ne$ leading to a current scaling like $\sum \tau_i ^n $. Below the gap $|eV| < 2\Delta$, the current $I$ is thus significant only in junctions with well-transmitted conduction channels, being highly nonlinear in ${\tau_i}$, and the $I(V)$ characteristic displays kinks at voltages $eV = 2\Delta/n$ that are indicative of MAR.
Previous studies on superconducting atomic contacts~\cite{bretheau_superconducting_2012} and semiconducting nanowires~\cite{goffman_conduction_2017} have shown that fitting these features using MAR theory allows to estimate the number of conduction channels and their transmissions~\cite{scheer_conduction_1997, riquelme_distribution_2005}. We apply the same method to estimate the transmission distribution of conduction channels in our pinJJs.

\begin{figure}
\includegraphics[width=1\columnwidth]{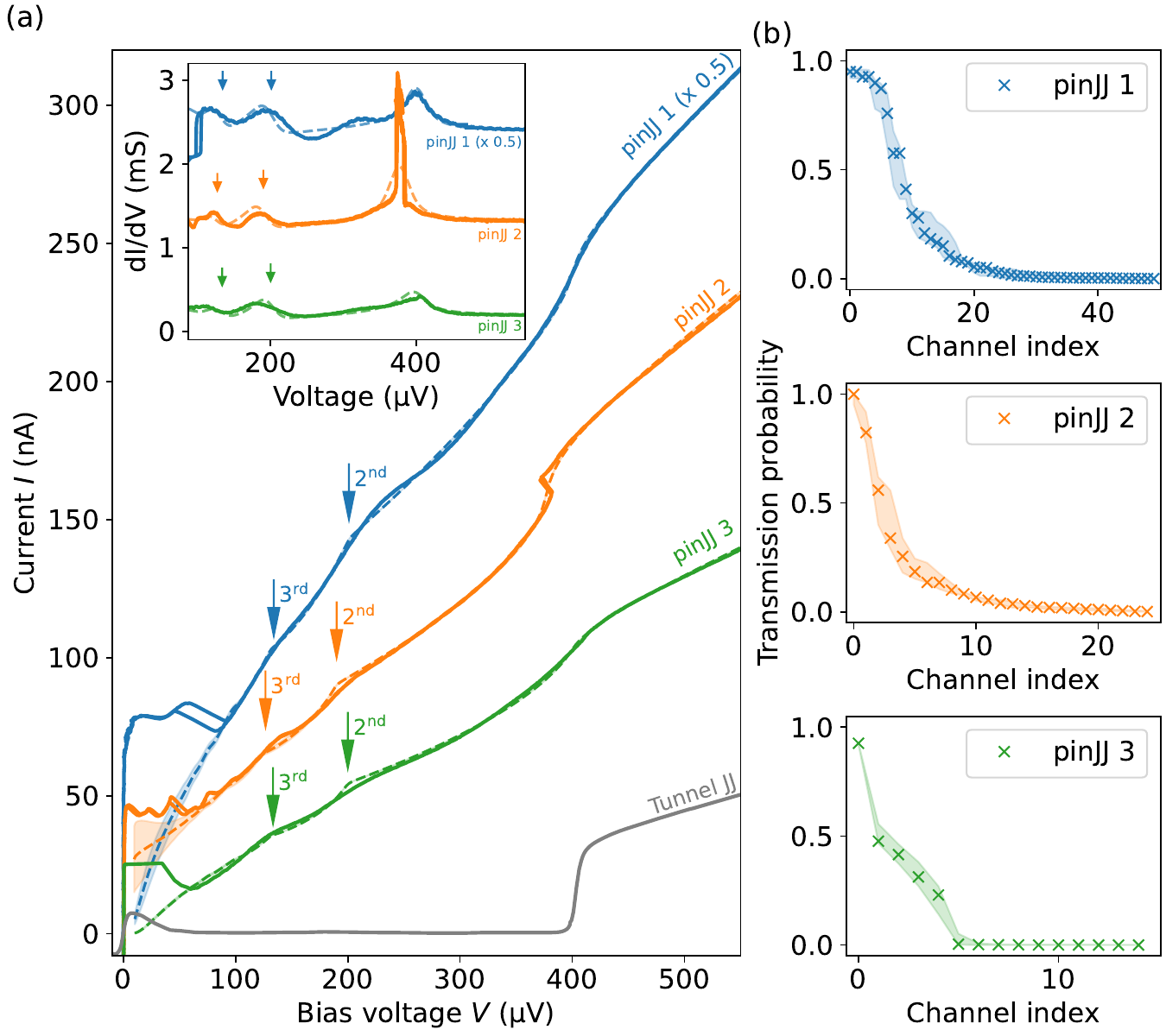}
\caption{\label{fig:MAR}MAR observation and channel characterization.
(a) Current-voltage characteristics of four Josephson junctions: pinJJs 1, 2 and 3 are pinhole junctions and tunnel JJ is a tunnel junction. The colored arrows highlight the kinks due to the MAR. The number next to them correspond to the MAR order. The dashed lines represent the best fits of the data. 
Inset: numerical differential conductance plot of pinJJs 1 and 2. Both data are filtered using a savitzky golay filter of order 1 in a \qty{20}{\micro\volt} window. The data for pinJJ 1 and 2 are shifted upwards by 2 and 1 mS for clarity. The dashed lines are the numerical derivatives of the best fits, convoluted with gaussians of width \qty{10}{\micro\volt} to account for thermal noise.
(b) Distribution of the conduction channel transmissions that best fits the data in (a).
In both panels the shaded areas represent the standard deviation of the different outcomes of 101 numerical fits. 
}
\end{figure}

Figure \ref{fig:MAR}a shows the current-voltage $I(V)$ characteristics measured at 10 mK for four pinJJs with resistances $R_J$ ranging from 1 to 10 k$\Omega$, illustrating four different distributions of conduction channels. The used setup is described in Appendix~\ref{sec:measurement}. Further data are available in Appendix~\ref{sec:SM_additionnalIV}. The junction with the largest resistance (grey curve) displays the $I(V)$ characteristic of a standard tunnel junction with three distinct parts: a supercurrent branch, an insulating behavior within the superconducting gap ($|eV| < 2\Delta$), and a quasiparticle branch at large voltage with a resistance close to $R_J$. The insulating region indicates that, for this particular junction, all conduction channels have very low transmission ($\tau \ll 1$), suggesting that no pinhole formed during the nanofabrication process.

The three other $I(V)$ curves show an additional subgap current due to MAR, recognizable by kinks at $eV = 2\Delta/n$. 
These features manifest more strongly as bumps in the numerical $dI/dV$ curves (see Inset).

By using the fitting procedure mentioned earlier~\cite{riquelme_distribution_2005}, we achieve a good agreement between our measurements and the MAR theory displayed as dashed lines in Fig. \ref{fig:MAR}a and its inset. This allows us to extract the number of channels and their transmissions, which are reported in Fig. \ref{fig:MAR}b. For low-resistance pinJJs  having $R_J\lesssim 1$~k$\Omega \ll R_Q$, we typically extract tens of channels with transmissions greater than 0.1, such as pinJJ 1 which has around 20 well-transmitted channels. For resistances approaching the quantum of resistance $R_J\sim R_Q$, like pinJJs 2 and 3 in Fig. \ref{fig:MAR}, the MAR current and the quasiparticle branch can be reproduced with distributions of only 5 to 10 well-transmitted channels with $\tau > 0.1$. The shaded areas in the figure show a better reliability of the fit in the case of junctions with fewer channels (see Appendix~\ref{sec:SM_fiterrorbar}).

Because our nanofabrication process does not permit fabricating electrodes with identical thicknesses and hence identical superconducting gaps, we leveraged the versatility of our platform to engineer asymmetric junctions combining two different superconducting materials: aluminum and titanium, which exhibit markedly distinct gaps. This configuration yields well-separated, sharply defined kinks. We performed analogous $I(V)$ measurements on these devices and use an extended model to account for the two-gap configuration. Beyond enabling clear identification of multiple Andreev reflection features, such junctions may also serve as useful platforms for probing emergent superconducting phenomena, including the Higgs mode~\cite{lahiri_ac_2024}. The enhanced gap contrast allows us additionally to clearly track the temperature evolution of these kinks and to observe their characteristic BCS-like behavior, as detailed in Appendix~\ref{sec:SM_additionnalIV} and \ref{sec:SM_Tdep}.

It is remarkable that, despite the presence of millions of tunneling channels in parallel with the small set given by the fit, the latter are sufficient to account for all transport properties, including the critical current value, the MAR current, and the junction's resistance on the quasiparticle branch. Therefore, we can reasonably expect that these tunneling channels will have a negligible impact on the current-phase relationship of the pinJJs and will not add a significant $\cos(\varphi)$ contribution to their Josephson potential, which would otherwise dilute the $\cos(n\varphi)$ higher-order harmonics.

\section{Estimation of higher harmonics}

\begin{figure}
\includegraphics[width=1\columnwidth]{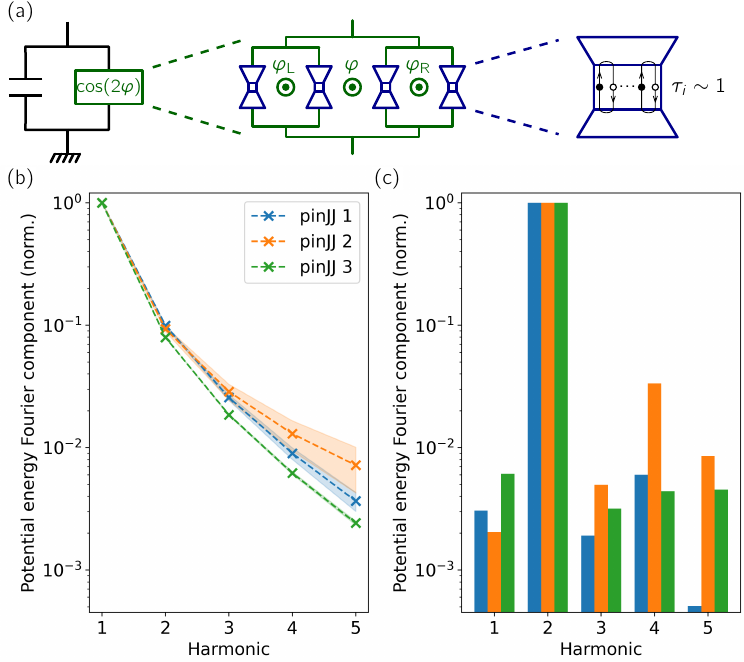}

\caption{\label{fig:cos2phi}Physical realization of a $\cos(2\varphi)$ qubit.
(a) From left to right: electric diagram for a $\cos(2\varphi)$ qubit including a $\cos(2\varphi)$ tunneling element and a capacitor. Physical implementation of the $\cos(2\varphi)$ element with four well-transmitted Josephson junctions. One Josephson junction hosts several conduction channels with transmission of the order of one.
(b) Absolute value of the Fourier coefficients of the potential energy corresponding to the channel distributions of Fig.~\ref{fig:MAR} (normalized by the 1st harmonic). The shaded areas represent the different outcomes of 101 numerical fits.
(c) Absolute value of the Fourier components of the potential energy for the four-junction circuit, for the flux values minimizing the non-$\cos(2\varphi)$ terms.
}
\end{figure}

Going further, we study the pinJJ's promise for generating higher harmonics content. Within the short well-coupled weak link approximation~\cite{beenakker_josephson_1991}, the Josephson potential is given by
\begin{equation}
\label{eq:ABS}
U_J(\varphi)=-\Delta\sum_i\sqrt{1-\tau_i\sin^2\left(\frac{\varphi}{2}\right)},
\end{equation}
where the sum runs over $m$ conduction channels of transmissions $\{\tau_1,\tau_2...\tau_m\}$. For $\tau_i\ll1$, we recover the cosinusoidal potential $-E_J \cos(\varphi)$ of a tunnel Josephson junction, with $E_J=\frac{\Delta}{4} \sum_i\tau_i$, which does not display any higher harmonics beyond $n=1$.
For pinJJs with well-transmitted channels, the Josephson potential acquires higher harmonics, which can be extracted through its Fourier decomposition. It can be expressed as $U_J=-\sum_n c_n \cos (n\varphi)$, where $n$ is a positive integer and $c_n$ are the Fourier coefficients. As shown in Fig. \ref{fig:cos2phi}b, these coefficients, computed from the transmission distributions of our pinJJs, tend to decrease with $n$. The second harmonic is typically an order of magnitude smaller than the first one, which leave the Josephson potential slightly modified compared to a tunnel junction. Nevertheless, these harmonics remain significant enough to be exploited. This requires to isolate one harmonic, which can be done by canceling the others through destructive interference in a multi-SQUID geometry. 

As an illustration, we propose a device, shown in Fig. \ref{fig:cos2phi}a, where destructive interference can be exploited to engineer a $\cos(2\varphi)$ element, \emph{i.e.} an element with only a $\cos(2\varphi)$ harmonic. As in recent publications~\cite{larsen_parity-protected_2020, schrade_protected_2022, leblanc_gate-_2025}, the idea is to have two tunable junctions in parallel. In our case, this is achieved by using two SQUIDs in parallel, made with two pinJJs each. This forms in total three loops that enclose reduced magnetic fluxes $\varphi_L$, $\varphi_R$, and $\varphi$, each of which can be individually controlled using local flux lines. By carefully adjusting these fluxes, it is then possible to cancel out the Fourier coefficients of the $n$-th harmonic, \emph{i.e.} $\sum_{k=1...4} c_{kn} e^{in\varphi_k} = 0$, where $k$ labels the four junctions, $c_{kn}$ is the $n$-th Fourier coefficient of junction $k$ and we define $\varphi_1=0$, $\varphi_2=\varphi_L$, $\varphi_3-\varphi_2=\varphi$, $\varphi_4-\varphi_3=\varphi_R$. For the particular case of four identical junctions, the first harmonic can be suppressed by choosing $\varphi_3=\pi$ and $\varphi_4=\varphi_2+\pi$ for any choice of $\varphi_2$. A remarkable property of such a perfectly symmetric device is that, for this choice of fluxes, all odd harmonics are suppressed, making it ideal to realize a parity protected qubit. In practice, the four junctions are nonetheless never identical, and odd harmonics might not entirely cancel.

In Fig. \ref{fig:cos2phi}c, we present the Fourier coefficients of Josephson potentials computed for various sets of four non-identical pinJJs in parallel. The fluxes have been optimized to achieve a potential that closely resembles $\cos(2\varphi)$. To account for the randomness in transmission distributions, these simulations were performed with quadruplets of pinJJs that are similar but not perfectly identical. Starting with the distributions obtained in Fig. \ref{fig:MAR}, we randomly modified the transmission of each channel by $\delta\tau\in[-0.1,0.1]$ independently for each junction (procedure details in Appendix~\ref{sec:cos2phiopt}). Even in this realistic configuration, most harmonics except for $n=2$ are heavily suppressed, particularly the first and third ones. In Appendix~\ref{sec:cos2phiopt}, we go even further and consider an unrealistic circuit with very different transmission distributions. Although it is not possible to completely eliminate all odd harmonics in this imperfectly symmetric case, the ratios between the second and the other harmonics are reduced by about two orders of magnitude, slightly better than the 95~\% of second-harmonic purity recently measured in a Ge system~\cite{leblanc_gate-_2025}. These simulation results strongly suggest that pinJJs could play a crucial role in engineering $\cos(2\varphi)$ elements and consequently in creating parity-protected qubits on a large scale.

\section{Conclusion}
Using DC measurements, we demonstrate that our new nanofabrication method can produce tunnel junctions with a small surface area, whose transport properties are dominated by a few well-transmitted conduction channels. These pinJJs represent a promising resource for implementing protected qubits such as the parity protected $\cos(2\varphi)$ qubits~\cite{ioffe_possible_2002, doucot_pairing_2002, gladchenko_superconducting_2009, bell_protected_2014, smith_superconducting_2020, larsen_parity-protected_2020, schrade_protected_2022} or the recently proposed harmonium qubits~\cite{hays_non-degenerate_2025}. One of the major advantages of pinJJs is the ability to fabricate them in large quantities - hundreds or thousands - in a standard research cleanroom. Unlike S-N-S type junctions, where N would be a normal metal, pinJJs are made with an extremely short weak link, on the order of the tunnel barrier thickness. As a result, they are protected by the superconducting gap of aluminum rather than a minigap, making them suitable for high-frequency applications in the range of 1-100 GHz, such as high-frequency qubits~\cite{anferov_superconducting_2024} or Josephson emitters of classical~\cite{cassidy_demonstration_2017, yan_low-noise_2021, peugeot_two-tone_2024} and non-classical~\cite{rolland_antibunched_2019, peugeot_generating_2021} microwave light based on dynamical Coulomb blockade. PinJJs also open up new perspectives for the realization of nonreciprocal devices, such as Josephson diodes~\cite{souto_josephson_2022} or hybrid Josephson rhombi~\cite{banszerus_hybrid_2025}.

On top of that, unlike hybrid junctions that rely on semiconducting materials, we anticipate that pinJJs are weakly sensitive to disorder and charge noise, making them a viable building block for developing novel quantum technologies.

From a more fundamental perspective, pinJJs can provide junctions with an extremely limited number of channels with near-unity transmission, enabling the isolation of a pair of Andreev states to be used as fermionic qubits~\cite{zazunov_andreev_2003, janvier_coherent_2015}. The relative ease of fabricating a large number of pinJJs on the same chip opens up the achievable prospect of realizing Andreev qubits or superconducting spin qubits. Additionally, pinJJs seem to be a promising building block for creating Andreev molecule or polymer-type devices to explore the non-locality of the generated fermionic wavefunctions~\cite{pillet_nonlocal_2019, pillet_josephson_2023, matsuo_observation_2022, haxell_demonstration_2023}.

\begin{acknowledgments}
We thank H. Duprez for valuable discussions and his critical reading of the manuscript.
We also want to thank J.-L. Smirr and the SPEC of CEA-Saclay, in particular the Quantronics group, for their help on nanofabrication and microwave expertise.
JG acknowledges support of Agence Nationale de la Recherche through grant ANR-24-CE47-4077.
JDP acknowledges support of Agence Nationale de la Recherche through grant ANR-20-CE47-0003.
LB acknowledges support of the European Research Council (ERC) under the European Union's Horizon 2020 research and innovation programme (grant agreement No. 947707).
This work has been supported by the French ANR-22-PETQ-0003 Grant under the France 2030 plan.
ALY acknowledges support from grant TED2021-130292B-C41 and the ``Mar\'{i}a de Maeztu'' Programme for Units of Excellence in R\&D CEX2023-001316-M.
\end{acknowledgments}

\begin{appendix}

\section{Details of nanofabrication}
\label{detailfab}

\begin{figure}
\includegraphics[width=0.95\columnwidth]{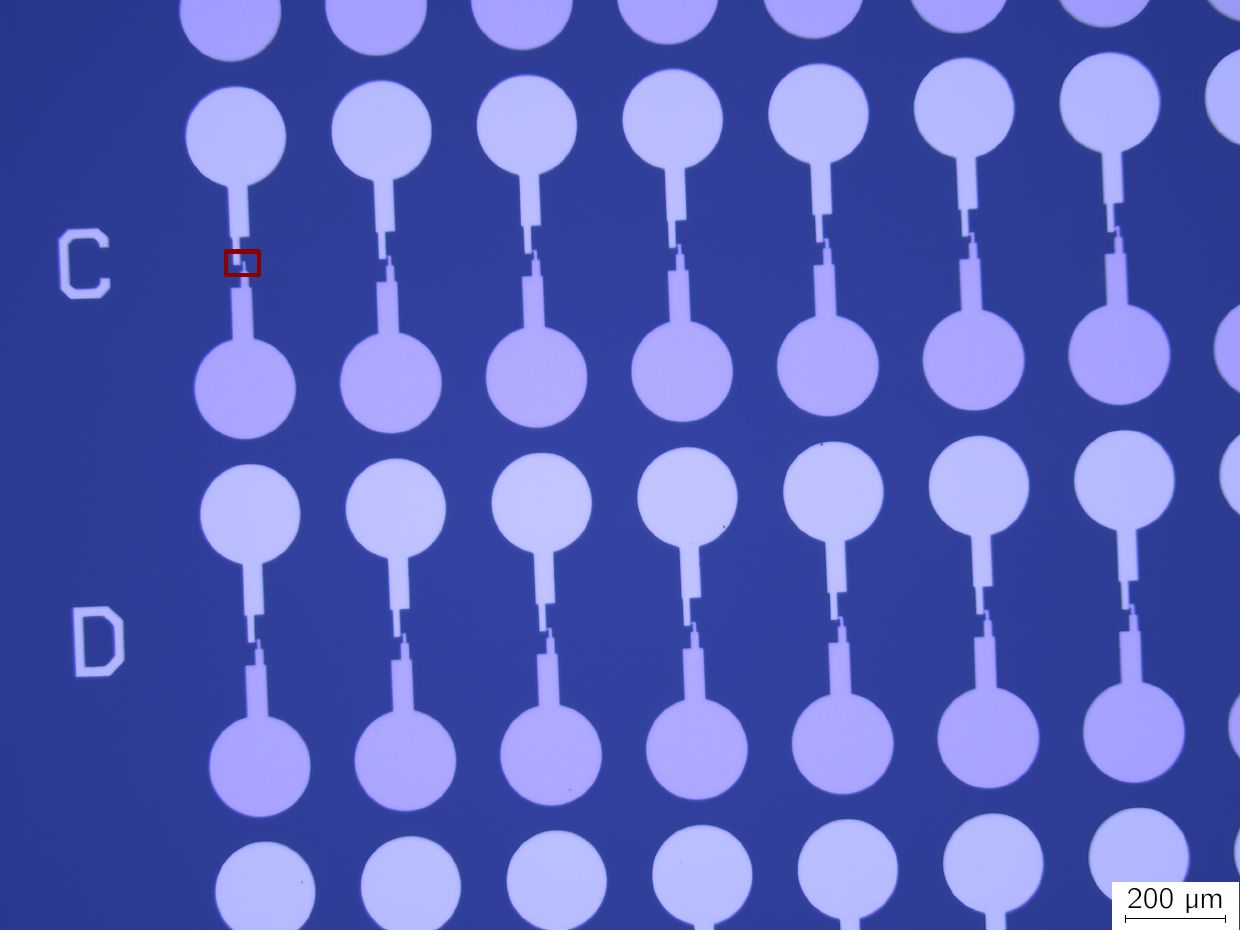}
\caption{\label{fig:SM_alotofJJs} Large scale microscope picture of a chip containing more than 50 pinJJs. The red box highlights the location of one junction.}
\end{figure}

We start from a silicon substrate, on which we pattern with a laser lithography an aluminum wire of typical cross section \qty{20}{\nano\meter}$\times$\qty{1}{\micro\meter} (Fig. \ref{fig:fab}a). 
A thick $\sim$\qty{1.5}{\micro\meter}-layer of optical resist (S1813) is then spun and a window is patterned for the design of the second electrode.
The first aluminum layer is scrapped with a strong argon milling process at an angle of \ang{10} with respect to the vertical axis (Fig. \ref{fig:fab}b). 
The emission current (30 mA) and milling time (10 min) of this step are chosen sufficiently high to totally remove the aluminum and its oxide.
This process is expected to create a rough surface where the junction will be, as can be seen in Fig.~\ref{fig:fab}c.
The remaining aluminum is then oxidized in a controlled oxygen environment.
The oxidation pressure and time are typically a few mbar and a few minutes.
Without breaking the vacuum, a second aluminum layer is evaporated with a \ang{20} angle with respect to the vertical, as visible in Fig.~\ref{fig:fab}d, such that a good overlap with the first layer is obtained.
For that purpose, the thickness of the second layer is also made much larger (\qty{100}{\nano\meter}) than the first one.
Fig. \ref{fig:SM_alotofJJs} shows a large scale microscope picture of a typical chip, containing a large quantity (>50) of pinJJs. The junctions (barely visible at this scale) are located in the center of each pair of pads (red box for the sample in the top left of the image).

\section{Room temperature characterization}
\label{sec:RT_R}
\begin{figure}
\includegraphics[width=0.95\columnwidth]{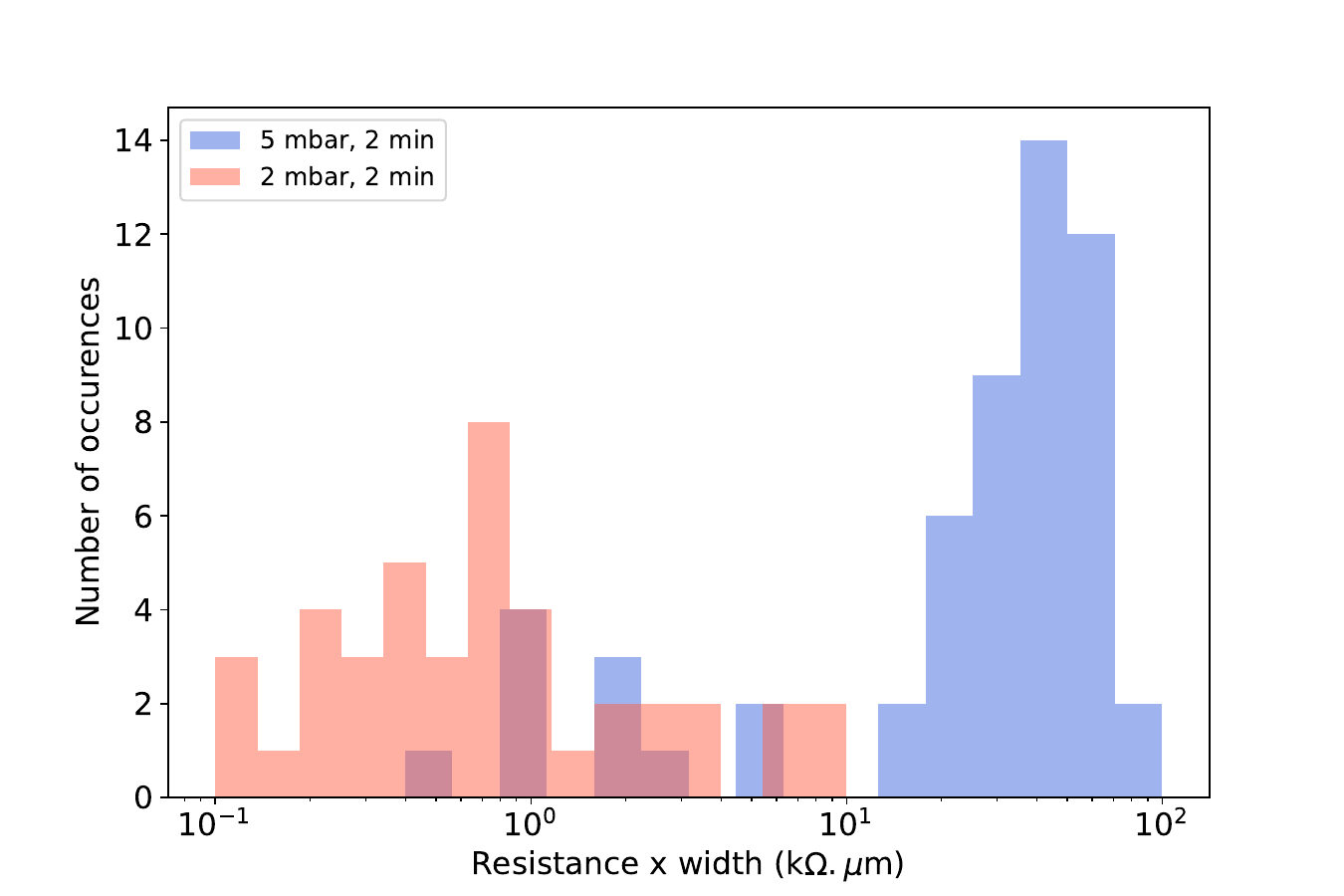}
\caption{\label{fig:SM_RT}Histogram of junction resistances measured on two chips fabricated with different oxidation parameters.}
\end{figure}

Fig.~\ref{fig:SM_RT} shows the typical resistance spread observed when fabricating multiple junctions on a single chip.
The broad distribution suggests that the transmissions of the conduction channels is randomly distributed, meaning that the formation of pinholes is a stochastic process. As expected, a reduced oxidation leads to a lower average resistance.

\section{Measurement setup}
\label{sec:measurement}

\begin{figure}
\includegraphics[width=0.95\columnwidth]{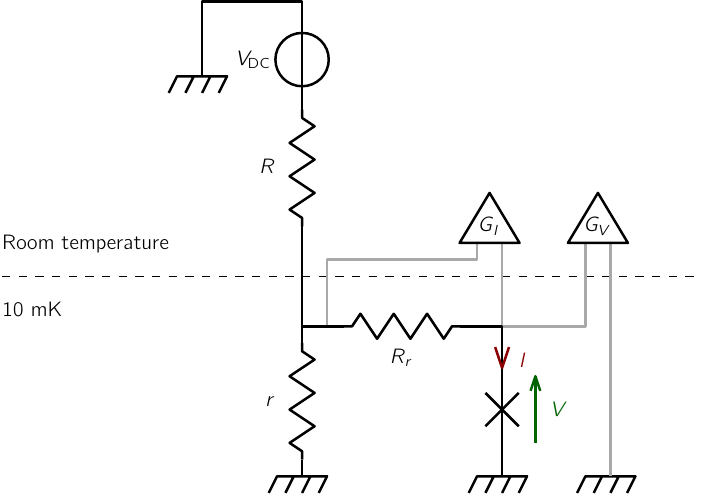}
\caption{\label{fig:SM_setup}Electrical diagram of the measurement setup. The junction is biased across a voltage divider placed at 10 mK. Both voltage $V$ and current $I$ are measured.}
\end{figure}

The sample (represented as a cross in Fig.~\ref{fig:SM_setup}) is thermally anchored to the bottom plate of a dilution cryostat at a temperature close to 10 mK. It is biased by a DC source (Yokogawa GS200) in series with a bias resistor $R=1~\mathrm{M}\Omega$ . This voltage is divided at low temperature by a small resistance $r=100~\Omega$. The reading resistor $R_r=100~\Omega$ in series with the junction allows to directly measure the current $I$ flowing through the junction.

The voltages across the reading resistor and the junction are amplified at room temperature by pre-amplifiers from Basel Precision Instruments (SP1004) with gains $G_I=10^4$ and $G_V=10^{4}$. The grey lines are the measurement lines and together with the feed line are filtered with home-made RC filters with a frequency cut-off around 1 kHz (not in Fig.~\ref{fig:SM_setup}). These lines are further filtered by Thermocoax cables.

The outputs of the amplifiers are connected to an oscilloscope (Yokogawa DL350) to measure current and voltage. The bias voltage is swept with a triangle shape at a frequency close to \qty{1}{\hertz} and the traces on the oscilloscope are taken with a small bandwidth of tens of \unit{\hertz}.

\section{Additional $I(V)$ curves}
\label{sec:SM_additionnalIV}

\begin{figure}
\includegraphics[width=0.95\columnwidth]{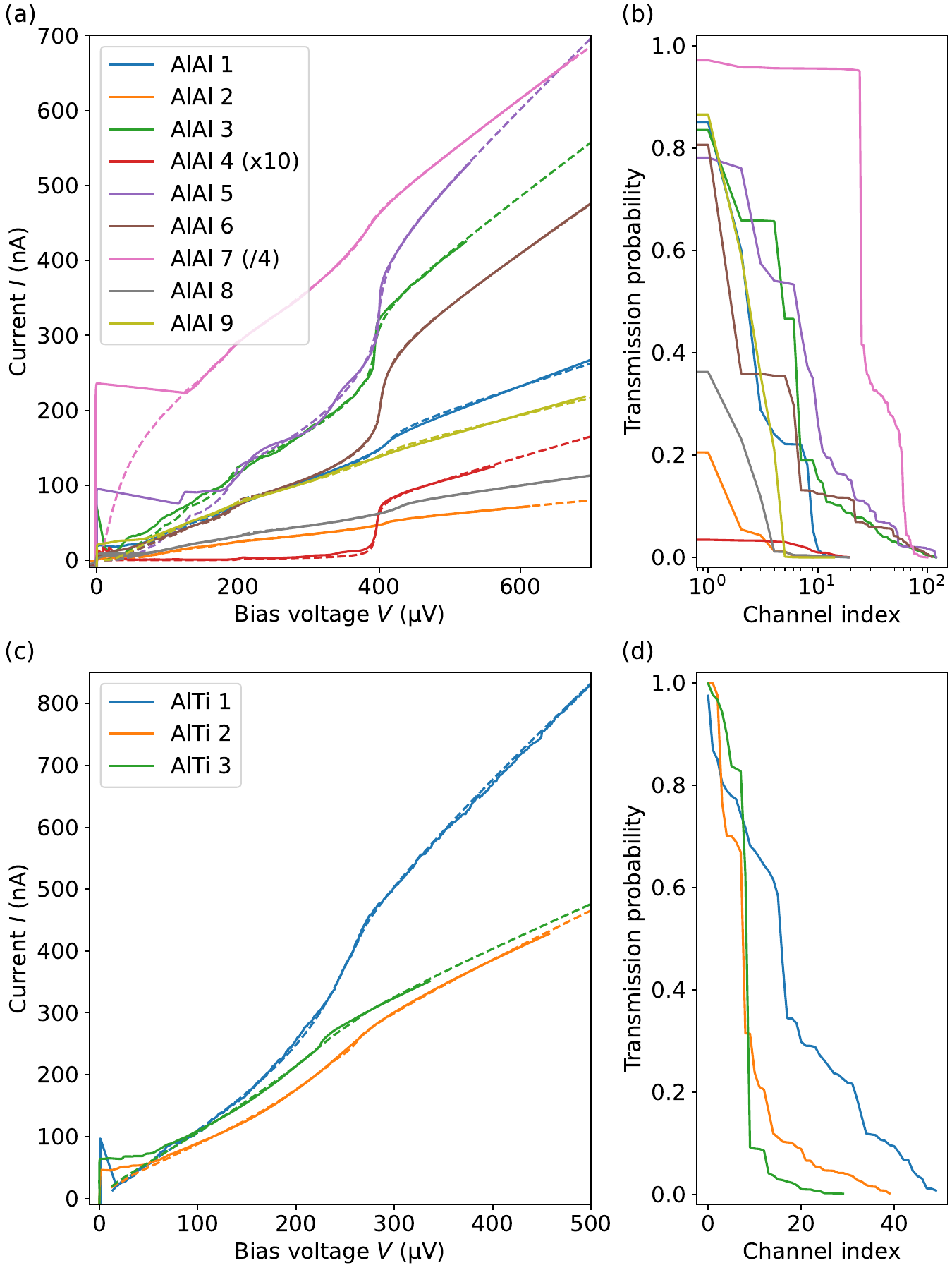}
\caption{\label{fig:SM_IVs} $I(V)$ curves exhibiting MAR for additional (a) Al/Al junctions and (c) Al/Ti junctions. (b) and (d) corresponding channel distributions that best fit the data in (a) and (c).}
\end{figure}

\begin{figure}
\includegraphics[width=0.95\columnwidth]{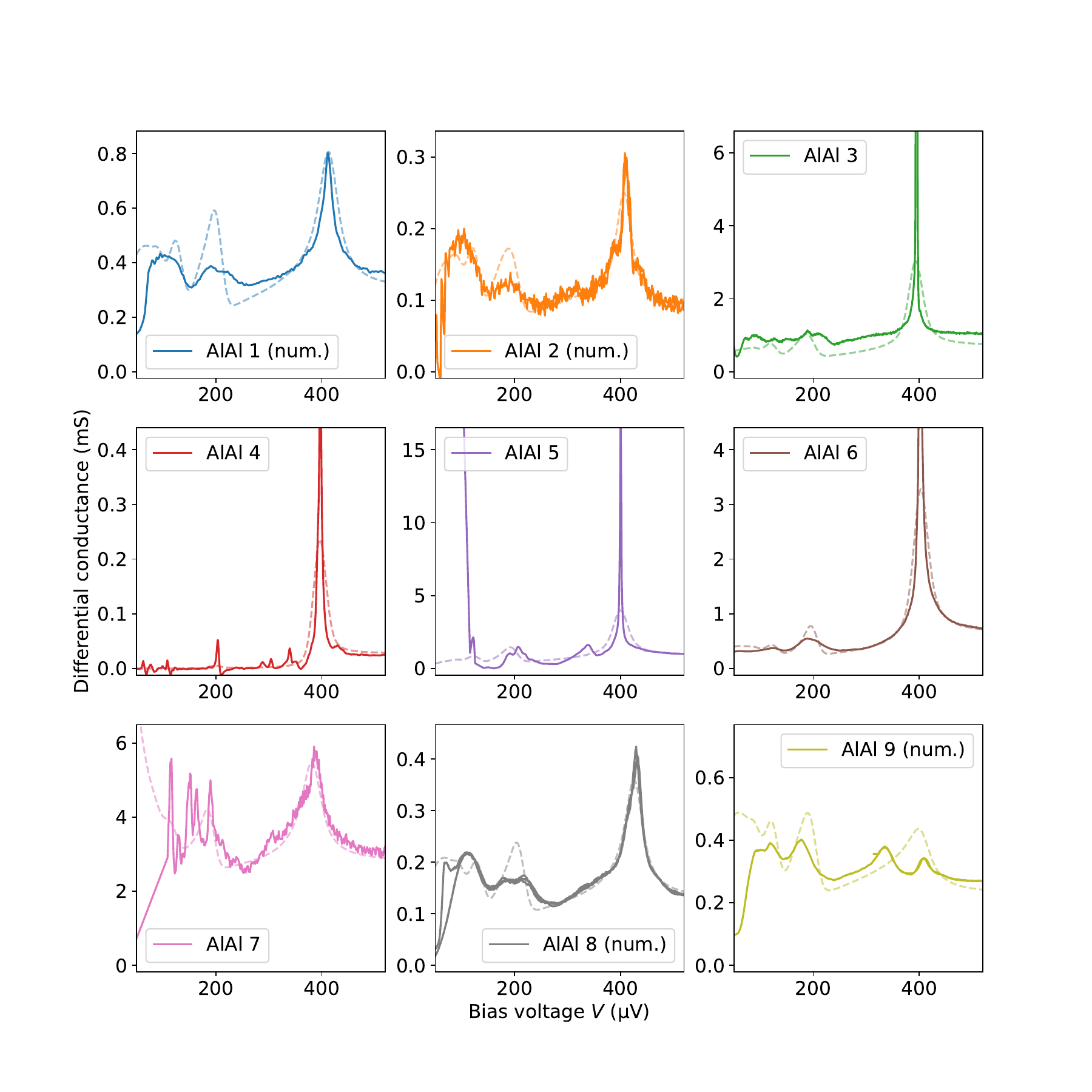}
\caption{\label{fig:SM_dIdVs} Differential conductance curves for additional Al/Al junctions. The names correspond to the $I(V)$ curves shown in Fig.~\ref{fig:SM_IVs}a. Those including (num.) in their name are numerical derivative and those without are taken with a lockin. The dashed lines are numerical derivative of the best fits shown in Fig.~\ref{fig:SM_IVs}a.}
\end{figure}

In addition to the samples whose $I(V)$ curves are shown in Fig.~\ref{fig:MAR}, we fabricated and measured several other pinJJs. Most of them were fabricated using the recipe described in Appendix~\ref{detailfab}. We plot the $I(V)$ curves of 9 of them in Fig.~\ref{fig:SM_IVs}a. As in the main text, we numerically fit them (dashed lines) using the same model as before and obtain the channel distributions shown in Fig.~\ref{fig:SM_IVs}b. They are representative of the diversity of behaviors we can obtain with this fabrication method: from almost tunnel (AlAl 4 in red) to almost transparent (AlAl 7 in pink). The corresponding differential conductances are shown in Fig.~\ref{fig:SM_dIdVs}. They follow qualitatively (and almost quantitatively) the curves obtained from the numerical fits.

We also fabricated asymmetric junctions, where the second layer of aluminum is replaced by titanium which has a smaller critical temperature and thus superconducting gap. These junctions are labelled AlTi $1-3$ and their $I(V)$ curves are plotted in Fig.~\ref{fig:SM_IVs}c. We extended the model used for symmetric junctions~\cite{riquelme_distribution_2005} in order to include two different gaps and obtain the fits plotted in dashed lines in the figure. The corresponding channel distributions are visible in Fig.~\ref{fig:SM_IVs}d.

\section{Temperature dependence of the MAR}
\label{sec:SM_Tdep}

\begin{figure}
\includegraphics[width=0.95\columnwidth]{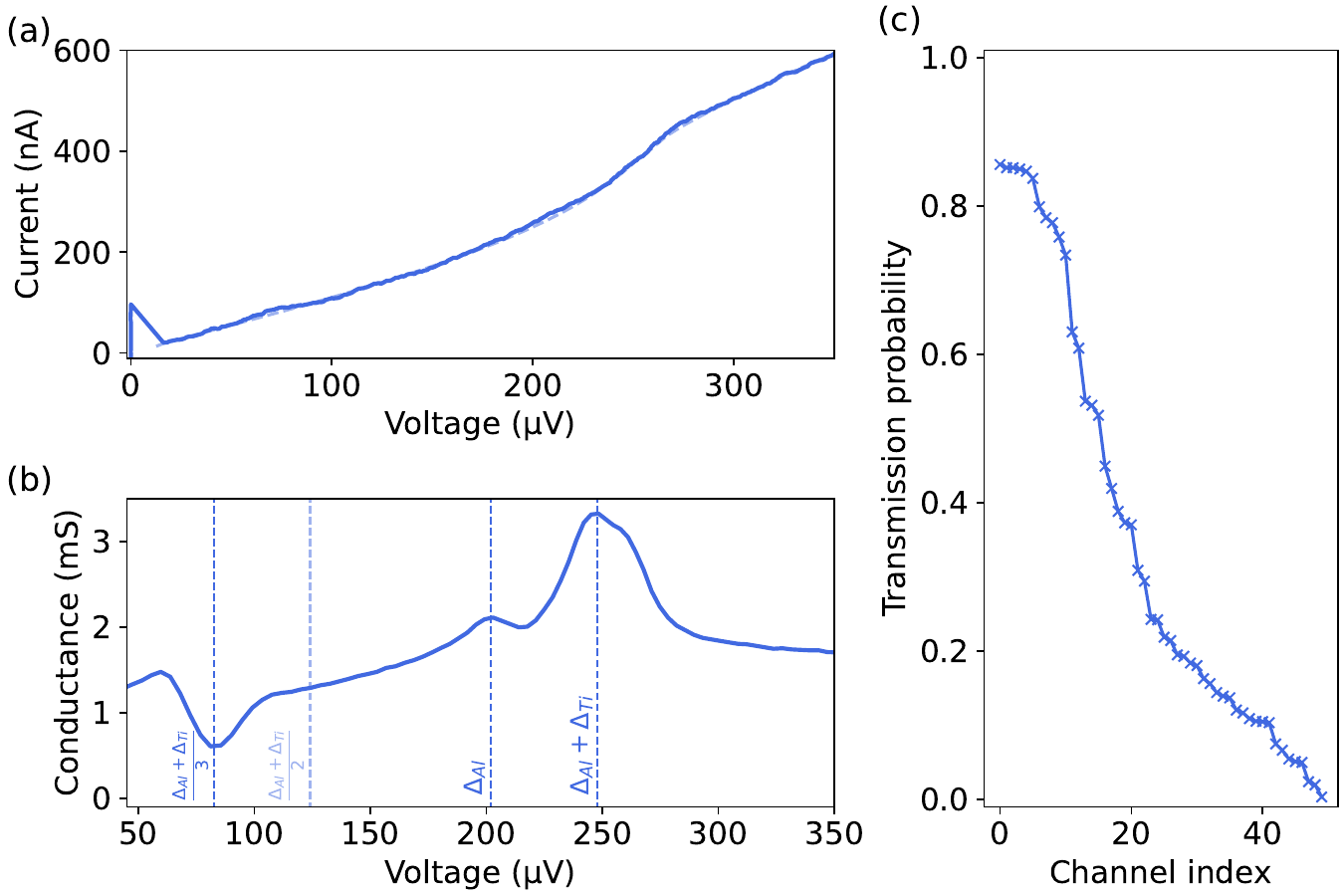}
\caption{\label{fig:SM_dIdV} $I(V)$ curve for an additional  Al/Ti junction: AlTi4.
(b) Differential conductance of the same sample. The vertical dashed lines represent the position of expected MAR features.
(c) Corresponding channel distributions that best fits the data.}
\end{figure}

\begin{figure}
\includegraphics[width=0.95\columnwidth]{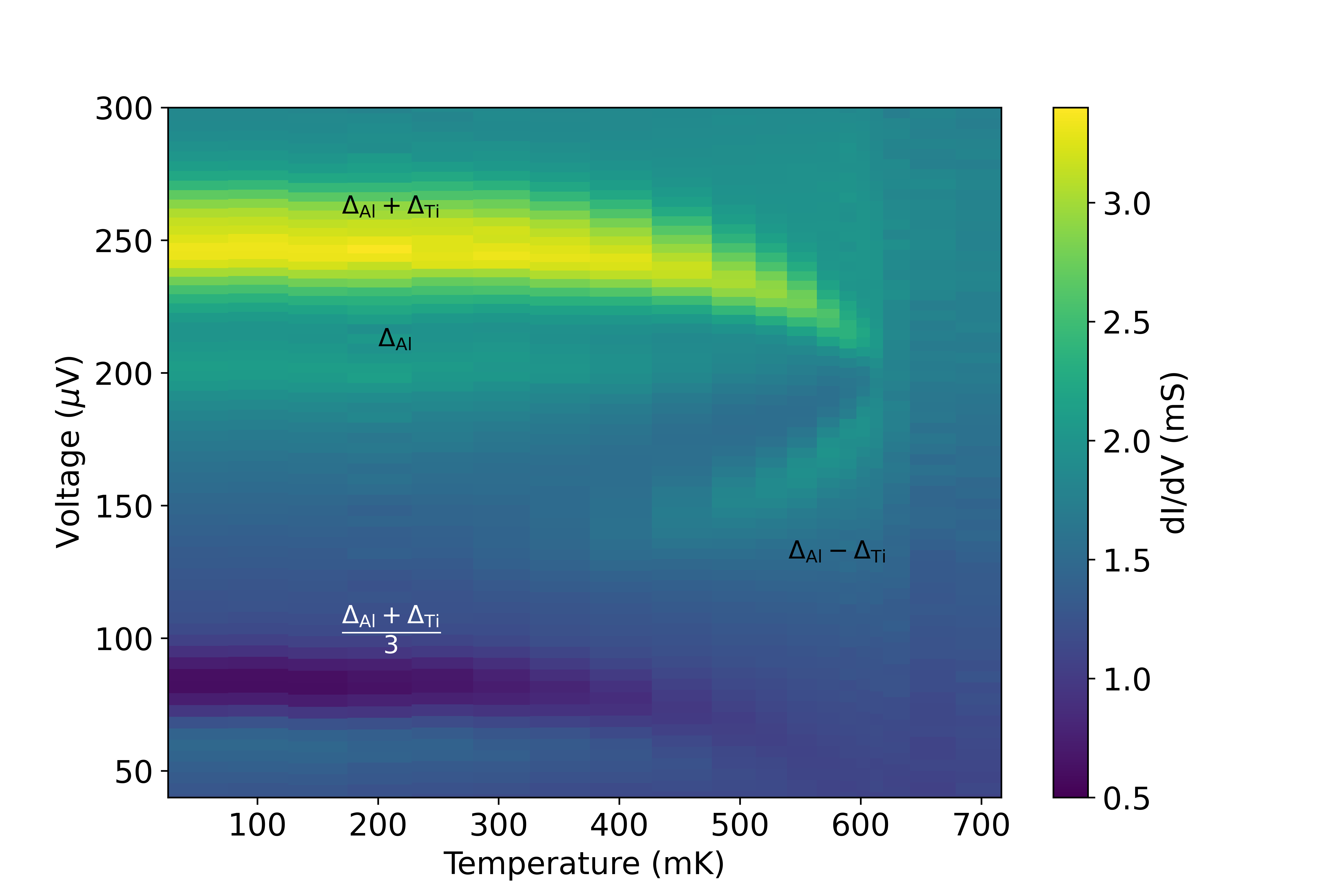}
\caption{\label{fig:SM_Tdep} Temperature dependence of the differential conductance of sample AlTi4.}
\end{figure}

We investigated in greater detail one asymmetric junction composed of aluminum and titanium. In contrast to the data shown in the inset of Fig.~\ref{fig:MAR}a, the differential conductance plotted in Fig.~\ref{fig:SM_dIdV}b was measured directly using a lock-in amplifier, rather than obtained through numerical differentiation. Distinct MAR features arising from the two different superconducting gaps are visible, with the most prominent peak occurring at the sum \(\Delta_\mathrm{Al} + \Delta_\mathrm{Ti}\).
Figure~\ref{fig:SM_Tdep} shows the temperature evolution of this differential conductance. As the temperature approaches the critical temperature of titanium (\(\sim\)\qty{650}{\milli\kelvin}), the peak at \(\Delta_\mathrm{Al} + \Delta_\mathrm{Ti}\) gradually shifts toward the peak at \(\Delta_\mathrm{Al}\), and the dip at \((\Delta_\mathrm{Al} + \Delta_\mathrm{Ti})/3\) also moves to lower voltage.
In addition to these standard MAR-related features, a new peak emerges above \qty{400}{\milli\kelvin} at a voltage corresponding to \(\Delta_\mathrm{Al} - \Delta_\mathrm{Ti}\). This feature is attributed to thermal excitation of quasiparticles in the smaller-gap superconductor (titanium), which can then tunnel into the larger-gap superconductor (aluminum).

\section{Error bars for the fit}
\label{sec:SM_fiterrorbar}

\begin{figure}
\includegraphics[width=0.95\columnwidth]{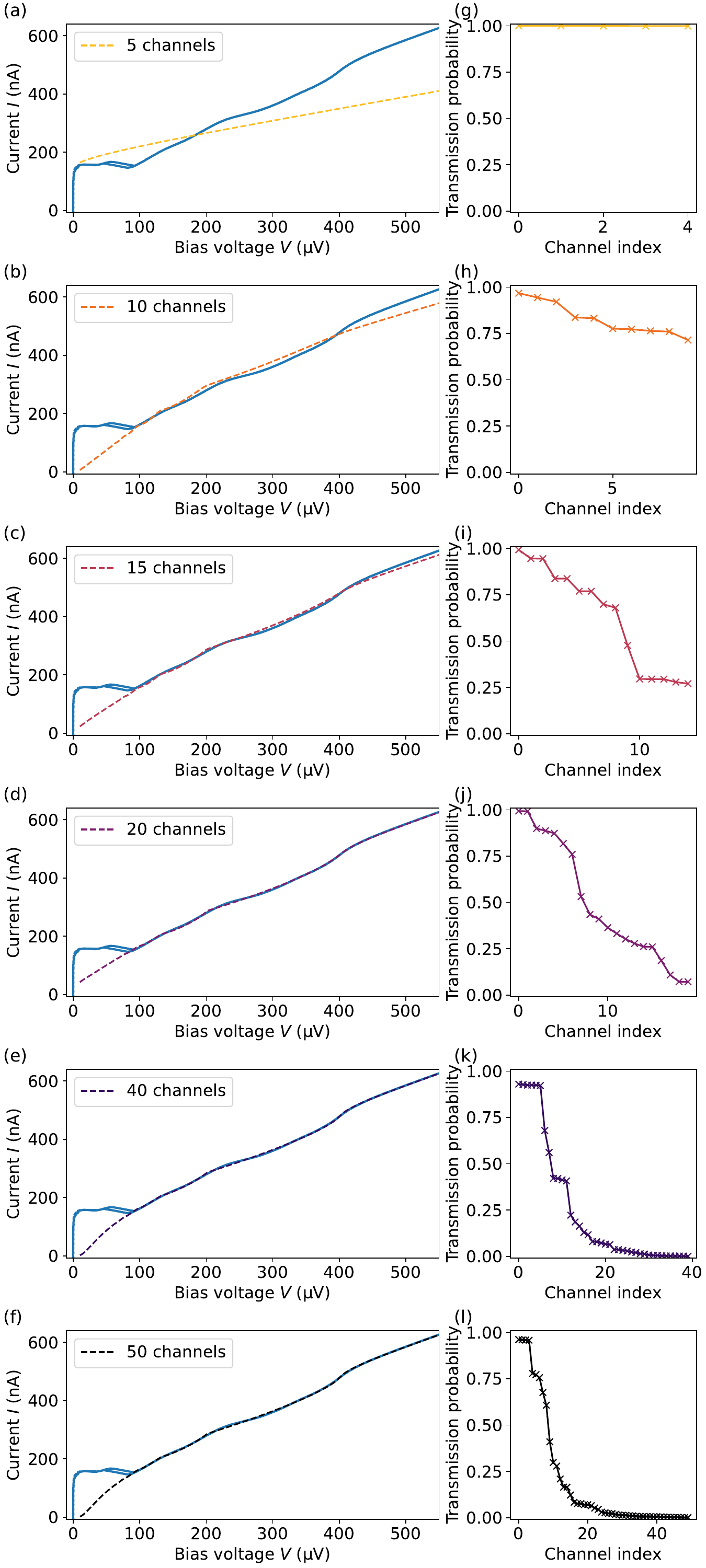}
\caption{\label{fig:SM_nch} (a)-(f) $I(V)$ curve for pinJJ 1 in blue. The dashed curves correspond to the best fit of the data with different number of channels, from 5 channels in (a) to 50 channels in (k). (g)-(l) Corresponding channel distributions.}
\end{figure}

To determine the appropriate number of channels for the fit, we perform the fitting procedure with an increasing number of channels and stop when the resulting channel distributions begin to converge.
Fig.~\ref{fig:SM_nch} illustrates this process for pinJJ 1.
With 5 channels (yellow curve), the fit in panel (a) fails to converge, indicating that this number is insufficient to account for the measured conductance. This is even more apparent in the transmission distribution plot, where all five channels are assigned a transmission probability of 1. As the number of channels increases (10, 15 and 20), the fit progressively improves. However, the smallest transmission probability remains greater than zero, suggesting that additional channels are still required.
When using 40 and 50 channels, the resulting distributions become nearly identical. At this point, the fitting algorithm primarily assigns zero transmission probabilities to the last 10 channels, indicating convergence.

To quantify the reliability of the distribution obtained with the fit, we repeated the fitting procedure 101 times for each junction. For each iteration, we started from a random distribution for the channels and used the \texttt{optimize.curve\_fit} function from the \texttt{scipy} package~\cite{scipy} to find the best channel distribution. 

The variation in the channel distributions is almost unnoticeable in the $I(V)$ curves (Fig. \ref{fig:MAR}a), except in the low voltage region, where the junction dynamics is anyhow dominated by switching processes and resonances in the electromagnetic environment~\cite{griesmar_superconducting_2021}.
In Fig. \ref{fig:MAR}b however, the width of the shaded areas indeed reflects the notable uncertainty of our fitting procedure for larger numbers of channels. Nevertheless, it provides a qualitatively correct idea of the number of transmission channels playing a significant role at the microscopic level.

\section{Simulation of the $\cos(2\varphi)$ element}
\label{sec:cos2phiopt}
To realize a $\cos(2\varphi)$ element, we consider the circuit shown in the central panel of Fig. \ref{fig:cos2phi}a.
For each of the three pinJJs considered in the article, we can write the total potential energy of the circuit as
$$U(\varphi)=\sum_{k=1}^{4}\sum_{n>0}c_{kn}\cos\left( n\left(\varphi-\varphi_k\right)\right),$$
where the $c_{kn}$ are the Fourier coefficients of the $n$-th harmonic of junction $k$ in the circuit. We define $\varphi_1=0$, $\varphi_2=\varphi_L$, $\varphi_3-\varphi_2=\varphi$, $\varphi_4-\varphi_3=\varphi_R$.
For the simulation, we fix the $c_{1n}$ coefficients (of junction 1 in the circuit) to be the ones plotted in Fig. \ref{fig:cos2phi}b.
They correspond to the potential $U_1(\varphi)$ of the junction considered, with the fitted set of transmissions $\left\{\tau_i\right\}_{i>0}$.
To obtain the other $c_{kn}$ (with $k>1$), we start from this set and to each $\tau_i$, add a random $\delta\tau\in[-0.1,0.1]$ (and force the new $\tau_i$ to stay between 0 and 1).
Using formula \eqref{eq:ABS}, we obtain potentials $U_2(\varphi)$, $U_2(\varphi)$ and $U_3(\varphi)$ for the three other junctions of the circuit.
We then calculate the Fourier coefficients of these potentials and obtain all the $c_{kn}$.

To obtain the potential $U(\varphi)$ of the total circuit closest to a pure $\cos(2\varphi)$, we numerically fit $U(\varphi, \varphi_1, \varphi_2, \varphi_3)$ to a $\cos(2\varphi)$ function with free amplitude, offset and phase origin for a given $(\varphi_1, \varphi_2, \varphi_3)$ triplet.
The covariance of this fit is then fed into the \texttt{optimize.minimize} function from the \texttt{scipy} package~\cite{scipy} which finds the $(\varphi_1, \varphi_2, \varphi_3)$ triplet returning the best $U(\varphi)$ potential. 
This procedure is repeated 100 times and Fig. \ref{fig:cos2phi}c shows the Fourier coefficients of the realization with the smallest non-$\cos(2\varphi)$ terms.

\begin{figure}[h]
\includegraphics[width=1\columnwidth]{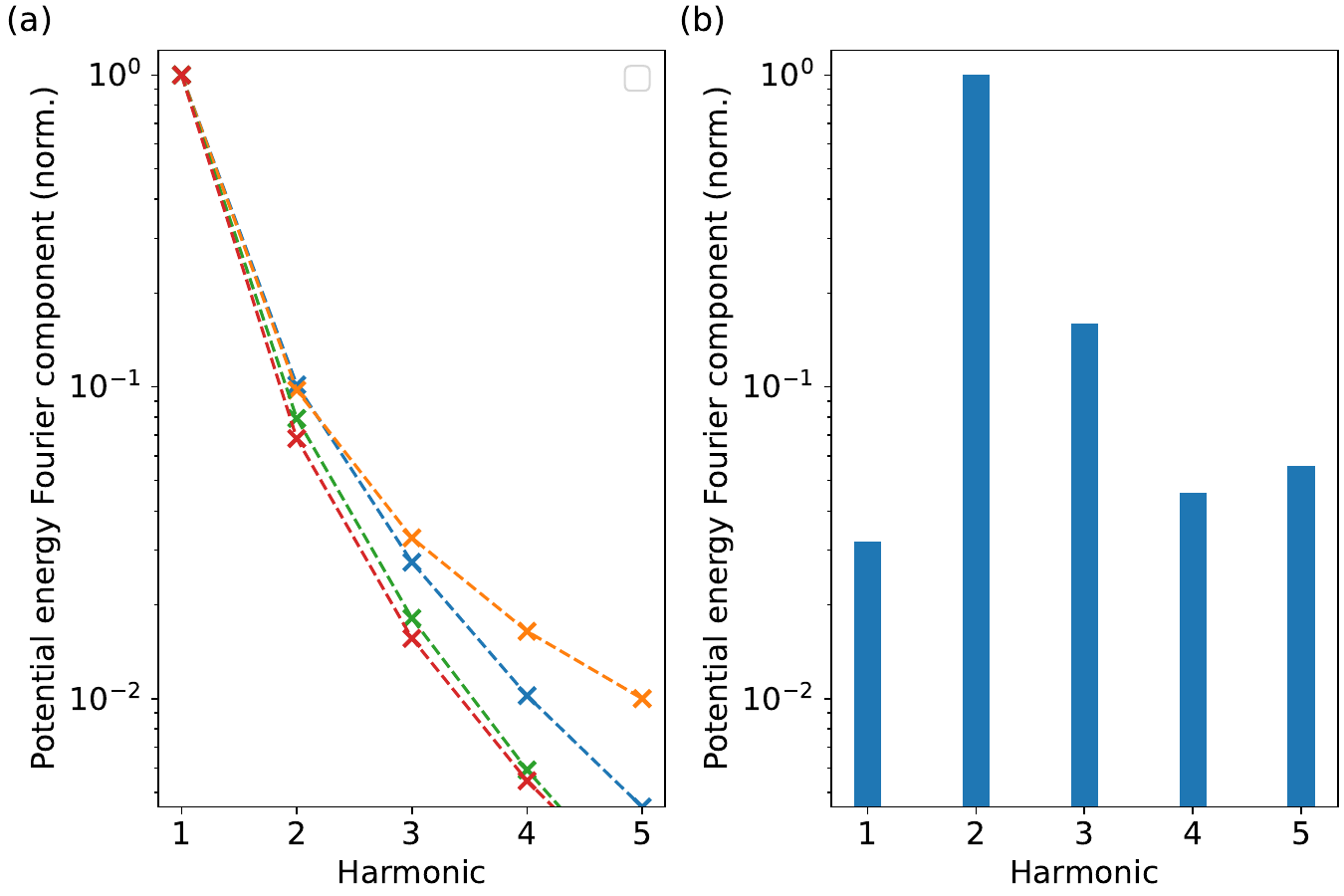}

\caption{\label{fig:supp_Fourier}(a) Absolute value of the Fourier coefficients of the potential energy corresponding to the channel distributions of Fig.~\ref{fig:MAR} (normalized by the 1st harmonic). 
(b) Absolute value of the Fourier components of the potential energy for the four-junction circuit, for the flux values minimizing the non-$\cos(2\varphi)$ terms.
}
\end{figure}

In Fig.~\ref{fig:supp_Fourier}, we simulated the potential that would result from a circuit made with the three junctions described in the main text (pinJJ 1, 2 and 3). The fourth junction transmissions are made with the average of the three other junctions transmission. Even in this scenario with junctions being extremely different, the 1st harmonic term becomes a few percents of the second harmonic term.

\end{appendix}

\bibliography{biblio}

\end{document}